\documentclass[fleqn,usenatbib]{mnras}

\usepackage{newtxtext,newtxmath}
\usepackage[T1]{fontenc}
\usepackage{ae,aecompl}

\usepackage{graphicx}	% Including figure files
\usepackage{amsmath}	% Advanced maths commands
\usepackage{amssymb}	% Extra maths symbols

\renewcommand{\d}{\mathrm{d}}	% upright differential

\title[Exploiting the hidden symmetry of Kerr black holes]{Exploiting the hidden symmetry of spinning black holes: conservation laws and numerical tests}

\author[Vojt{\v e}ch Witzany]{
Vojt{\v e}ch Witzany$^{1}$\thanks{E-mail: vojtech.witzany@zarm.uni-bremen.de}
\\
$^{1}$ZARM, University of Bremen, Am Fallturm 2, 28359 Bremen, Germany
}

\date{Accepted XXX. Received YYY; in original form ZZZ}

\pubyear{2015}

\begin{document}
\label{firstpage}
\pagerange{\pageref{firstpage}--\pageref{lastpage}}
\maketitle

\begin{abstract}
The Kerr black hole is stationary and axisymmetric, which leads to conservation of energy and azimuthal angular momentum along the orbits of free test particles in its vicinity, but also to conservation laws for the evolution of continuum matter fields. However, the Kerr space-time possesses an additional ``hidden symmetry" which exhibits itself in an unexpected conserved quantity along geodesics known as the Carter constant. We investigate the possibility of using this hidden symmetry to obtain conservation laws and other identities which could be used to test astrophysical simulations of the evolution of matter fields near spinning black holes. After deriving such identities, we set up a simple numerical toy model on which we demonstrate how they can detect the violations of evolution equations in a numerical simulation. Even though one of the expressions we derive is in the form of a conservation law, we end up recommending an equivalent but simpler expression that is not in the form of a conservation law for practical implementation.
\end{abstract}


%%%%%%%%%%%%%%%%%%%%%%%%%%%%%%%%%%%%%%%%%%%%%%%%%%

%%%%%%%%%%%%%%%%% BODY OF PAPER %%%%%%%%%%%%%%%%%%

\section{Introduction}\label{sec:intro}

The observational properties of black holes in systems of various scales from X-ray binaries to active galactic nuclei are computed from the behaviour of various test matter fields evolving in the Kerr space-time, the general-relativistic field of an isolated spinning black hole. For instance, in the case of accretion onto black holes the consensus has gradually emerged that one needs to include radiation, single-species hydrodynamics, and magnetic fields, but also possibly multi-species hydrodynamics or even non-Maxwellian rarefied plasma dynamics  to reproduce all the features of a real accretion process in a computer simulation  \citep{abramowiczlivrev, blaes2014}. 

With the increased complexity and additional layers of physics involved in such computer simulations, the question is how to test either for implementation mistakes or for inherent errors of the numerical evolution schemes. One particular way to do this is to see whether the codes reproduce the behaviour of analytical solutions to the dynamical equations \citep{harm,villiershawley2003}. However, there is only a very small set of solutions against which one can carry out such tests and they will typically probe only a subset of the implemented physics. Another approach, pursued for instance in \citet{harisconservation} or \citet{andersson}, is to find new conservation laws coming from the structure of the equations involved. We focus here on the latter approach, and specifically on formulating a conservation law or a similar expression which should be applicable to the evolution of any test matter field on the Kerr background, thus encompassing any possible model of the accretion process.

To do so, we investigate the possibility of a conservation law coming from the so-called ``hidden symmetry" of the Kerr field. The Kerr space-time is stationary and axisymmetric, which implies the conservation of orbital energy and azimuthal angular momentum along free test particle trajectories. The symmetries also lead to energy and angular moment currents that are conserved for any continuum test matter field evolving on the Kerr background. 

This, however, is not a full list of conservation laws in Kerr space-time; a nowadays classical analysis of \citet{carter1968} showed that there is an additional integral of motion for the free test particle motion, currently known as the Carter constant. The Carter constant is a square of an angular-momentum-like vector dragged along the trajectory and it cannot be linked to any explicit symmetry of the Kerr space-time, only to particular geometric properties of the Kerr metric such as the existence of a so-called Killing-Yano (KY) tensor \citep{walkerpenrose1970,floyd1973}. The existence of the KY tensor and other geometric structures in the Kerr space-time is exactly what is informally referred to as the ``hidden symmetry". 

One would expect a conservation law of a scalar quantity along single-particle trajectories to always have a direct counterpart in conservation laws for the evolution of continuum matter fields. However, in the case of the Carter constant and the hidden symmetry no conservation law for the evolution of general matter fields was known until now. An intuitive reason for this was given by \citet{grantflan2015}, who demonstrated that the sum of the Carter constants of a set of particles is not conserved once we allow for elastic collisions.

We find a loophole to this argument in Section \ref{sec:KYlaw} by studying instead the conservation of the sum of the angular-momentum-like vectors associated with the Carter constant, and obtain a conserved current indirectly associated with this conserved sum. Nonetheless, the resulting KY conservation law turns out to have a smaller potential for the detection of computation error than conservation laws coming from explicit symmetries. As a side product of our investigation, we realise that there exists an infinite family of conservation laws with properties similar to the KY conservation law, each of which provides a different ``basis'' in probing a possible violation of the equations of motion. Hence, this investigation could be in fact understood as a probe into this family of conservation laws by studying one representative member.

To assess more precisely the applicability of the KY conservation law, we devise a simple numerical demonstration in Section \ref{sec:demo}. We construct an analytical model of matter infalling into a black hole with a small unphysical acceleration in the direction of the meridional plane and numerically gauge several options of detecting such an acceleration from data given on a finite grid. Even though the KY conservation law can detect violations of the equations of motion undetectable by conventional means, we end up recommending an alternative method of detection based on the KY tensor that is not in the strict form of a conservation law but is much less demanding in terms of implementation and computational power. 

%%%%%%%%%%%%%%%%%%%%%%%%%%%%%%%%%%%%%%%%%%%%%%%%%%%%%%%%%%%%%%%%%%%%%%%%%%%%%%%%%%%%%%%%%%%%%%%%%%%%%%%%%%%%%%%%%%%%%%%%%%%%%%%%%%%%%%%%%%%%%%%%%%%%%%%%%%%%%%%%%
%%%%%%%%%%%%%%%%%%%%%%%%%%%%%%%%%%%%%%%%%%%%%%%%%%%%%%%%%%%%%%%%%%%%%%%%%%%%%%%%%%%%%%%%%%%%%%%%%%%%%%%%%%%%%%%%%%%%%%%%%%%%%%%%%%%%%%%%%%%%%%%%%%%%%%%%%%%%%%%%%

\section{Usual conservation laws} \label{sec:cons}
We use the $G=c=1$ geometrized units and the $-+++$ signature of the metric. We also assume that we have a 3+1 split of coordinates where $x^i$ are the coordinates on the spatial hypersurface and $t$ a temporal coordinate.

%%%%%%%%%%%%%%%%%%%%%%%%%%%%%%%%%%%%%%%%%%%%%%%%%%%%%%%%%%%%%%%%%%%%%%%%%%%%%%%%%%%%%%%%%%%%%%%%%%%%%%%%%%%%%%%%%%%%%%%%%%%%%%%%%%%%%%%%%%%%%%%%%%%%%%%%%%%%%%%%%

\subsection{Conservation of stress-energy}
We assume a general violation of the equations of motion which shows itself as a non-conservation of the stress-energy tensor $T^{\mu}_{\;\;\nu;\mu} = r_\nu$ where $r_\nu$ is some small non-physical residue induced either by an implementation mistake or numerical error. This equation can then be written as
\begin{equation}
\frac{1}{\sqrt{-g}}\left( T^{\mu}_{\;\; \nu} \sqrt{-g} \right)_{,\mu}  +  \frac{1}{2}g_{\alpha \beta, \nu} T^{\alpha \beta} = r_\nu\,. \label{eq:quasicons}
\end{equation}
Upon integration over a spatial volume $V$ with a surface $S$ from time $t_0$ to $t_1$ and the application of the divergence theorem we obtain
\begin{equation}
\begin{split}
&\left[ \int_V\! T^{t}_{\;\; \nu} \sqrt{-g} \,\d^3 x\right]_{t_0}^{t_1}  + \int_{t_0}^{t_1}\!\!\int_S \! T^{i}_{\;\; \nu}  \sqrt{-g} \,\d^2 S_i \d t 
\\
& + \int_{t_0}^{t_1}\!\!\int_V\! \frac{1}{2}g_{\alpha \beta, \nu} T^{\alpha \beta} \sqrt{-g} \,\d^3 x
 \, \d t  = \int_{t_0}^{t_1}\!\!\int_V\! r_\nu \sqrt{-g} \,\d^3 x \, \d t \,, \label{eq:quasiviol}
\end{split} 
\end{equation}
where $\d^2 S_i$ is the coordinate surface element. It is now possible to apply the integrated form of equations of motion to test for the presence of a residual on the right-hand side. However, this formula (with $r_\nu =0$) is often used directly in the construction of the so-called conservative numerical schemes (such as e.g. HARM, written by \citet{harm}) and testing the code against the integral identities (\ref{eq:quasiviol}) should indicate only coding mistakes, or computational errors which are already well understood. On the other hand, for {\em non}-conservative schemes such as that of \citet{villiershawley2003}, equation (\ref{eq:quasiviol}) can serve as an indirect test of the validity of the evolution.

%%%%%%%%%%%%%%%%%%%%%%%%%%%%%%%%%%%%%%%%%%%%%%%%%%%%%%%%%%%%%%%%%%%%%%%%%%%%%%%%%%%%%%%%%%%%%%%%%%%%%%%%%%%%%%%%%%%%%%%%%%%%%%%%%%%%%%%%%%%%%%%%%%%%%%%%%%%%%%%%%

\subsection{Conservation from explicit symmetries} \label{subsec:excons}
If the metric is independent of a coordinate $x^{\hat \nu}$ (such as time $t$ or azimuthal angle $\varphi$), we have $g_{\alpha \beta,\hat{\nu}} = 0$ and equation (\ref{eq:quasiviol}) simplifies as
\begin{equation}
\begin{split}
&\left[ \int_V\! T^{t}_{\;\; \hat \nu} \sqrt{-g} \,\d^3 x\right]_{t_0}^{t_1}  + \int_{t_0}^{t_1}\!\!\int_S \! T^{i}_{\;\;  \hat \nu}  \sqrt{-g} \,\d^2 S_i \d t 
\\
&  = \int_{t_0}^{t_1}\!\!\int_V\! r_{ \hat \nu} \sqrt{-g} \,\d^3 x \, \d t \,. \label{eq:expint}
\end{split}
\end{equation}
This makes the integral test much more powerful because one then needs to integrate only over a domain of dimension 3 to test for violations in a space-time volume of dimension 4. 

One particular demonstration of this power is to set the surface $S$ outside of an isolated system (so that $T^i_{\;\;\hat \nu}$ vanishes on $S$ at all times) and we then get a quantity $\int T^{t}_{\;\; \hat \nu} \sqrt{-g} \,\d^3 x$ which should be conserved at all times. This means that the noise coming from the numerical integration of the left-hand side of (\ref{eq:quasiviol}) becomes entirely independent of the time interval $[t_0,t_1]$ and we are able to search for an arbitrarily small average value of $r_{ \hat \nu}$ by letting its effect accumulate over very large times. 

A routinely implemented example is the conservation of total azimuthal angular momentum $\int T^{t}_{\;\; \varphi} \sqrt{-g} \,\d^3 x$ or energy $\int T^{t}_{\;\; t} \sqrt{-g} \,\d^3 x$ carried by an isolated matter field evolving in a stationary and axisymmetric space-time.  The inherent noise from the computation of total energy or angular momentum contained in the system is independent of the elapsed simulation time and one thus obtains a robust handle on cumulative error.

%%%%%%%%%%%%%%%%%%%%%%%%%%%%%%%%%%%%%%%%%%%%%%%%%%%%%%%%%%%%%%%%%%%%%%%%%%%%%%%%%%%%%%%%%%%%%%%%%%%%%%%%%%%%%%%%%%%%%%%%%%%%%%%%%%%%%%%%%%%%%%%%%%%%%%%%%%%%%%%%%
%%%%%%%%%%%%%%%%%%%%%%%%%%%%%%%%%%%%%%%%%%%%%%%%%%%%%%%%%%%%%%%%%%%%%%%%%%%%%%%%%%%%%%%%%%%%%%%%%%%%%%%%%%%%%%%%%%%%%%%%%%%%%%%%%%%%%%%%%%%%%%%%%%%%%%%%%%%%%%%%%

\section{KY conservation law} \label{sec:KYlaw}
Let us now turn our attention to the geometrical formulation of explicit symmetry. The fact that the metric does not change along a certain direction $\xi^\mu$ can be expressed by the fact that this vector fulfils the so-called Killing equation
\begin{equation}
\xi_{\mu;\nu} + \xi_{\nu;\mu} \equiv 2 \xi_{(\mu;\nu)} = 0\,. \label{eq:killing}
\end{equation}
In the set of coordinates where $x^{\hat{\nu}}$ is the symmetry coordinate the corresponding Killing vector has the components $\xi^\mu  = \delta^\mu_{\;\hat{\nu}}$. 

In this language, the integral of geodesic motion is written as $u^\mu \xi_\mu$
\begin{equation}
\frac{\d(u^\mu \xi_\mu)}{\d \tau} = u^\mu_{\;;\kappa} u^\kappa \xi_\mu + u^\mu u^\kappa \xi_{\mu;\kappa} = 0\,,
\end{equation}
where the first term vanishes due to the geodesic equation and the second term vanishes due to the Killing equation. The conservation law for continuum fields (\ref{eq:expint}) can then be understood as a consequence of the fact that $T^{\mu\nu}\xi_\nu$ is a conserved (divergenceless) current
\begin{equation}
\frac{1}{\sqrt{-g}}(T^{\mu\nu} \xi_\nu \sqrt{-g})_{,\mu} = (T^{\mu\nu} \xi_\nu)_{;\mu} = T^{\mu\nu}_{\;\;\;;\mu} \xi_\nu +  T^{\mu\nu} \xi_{\nu;\mu} = 0\,.
\end{equation}
One particular generalization of the Killing equation which is no longer connected to any explicit symmetry of the metric is the anti-symmetric KY tensor of rank two $Y_{\mu\nu} = - Y_{\nu \mu}$ which fulfils a direct generalization of the Killing equation of the form $Y_{\mu(\nu;\kappa)} =0$. One consequence of the existence of such a tensor is that geodesics parallel-transport a vector $L_\nu = u^\mu Y_{\mu\nu}$ along their motion
\begin{equation}
\frac{\mathrm{D}\,}{\d \tau}(L_\nu) = u^\mu_{\;;\kappa} u^\kappa Y_{\mu\nu}+  u^\mu Y_{\mu\nu;\kappa} u^\kappa = 0\,,
\end{equation}
where the first term again vanishes due to the geodesic equation and the second term vanishes due to the property $Y_{\mu(\nu;\kappa)} =0$. The most important consequence is that $L^\kappa L_\kappa$ is then an integral of motion.  In the case of the Kerr space-time, there exists a Killing-Yano tensor such that $L_\kappa$ represents a generalization of the angular momentum vector; $L^\kappa L_\kappa$ can then be understood as a generalized specific angular momentum squared and is known as the Carter constant \citep{floyd1973, penrose1973}. 

Our aim is to find a conserved current $j^\mu,\, j^\mu_{\;\,;\mu}=0$ associated with the hidden symmetry of the Kerr space-time so that we can obtain formulas similar to (\ref{eq:expint}).

%%%%%%%%%%%%%%%%%%%%%%%%%%%%%%%%%%%%%%%%%%%%%%%%%%%%%%%%%%%%%%%%%%%%%%%%%%%%%%%%%%%%%%%%%%%%%%%%%%%%%%%%%%%%%%%%%%%%%%%%%%%%%%%%%%%%%%%%%%%%%%%%%%%%%%%%%%%%%%%%%
\subsection{Physical motivation and statement of conservation law}

We now show the link between the integrals of free test particle motion and conservation laws for continuum fields. Furthermore, we clarify why certain integrals of motion have simple conservation-law counterparts while others do not. 

One could naively think that it is sufficient to find a conservation law fulfilled for a cloud of free-streaming particles due to the conservation laws along individual trajectories, and a conservation law for general continuum fields will then always follow. However, we demonstrate that such conservation laws can be spoiled already in the simple case of a cloud of free-streaming particles that are allowed to undergo an occasional elastic collision.

Let $p_{(i)}^{ \mu}$ be the momenta of the particles where $i$ is an index which runs over all the particles, we can then write that in an elastic collision
\begin{equation}
\sum_i p_{(i)}^{\prime \mu} - \sum_i p_{(i)}^{ \mu} = 0\,, \label{eq:momcons}
\end{equation}
where the primed and unprimed quantities always signify quantities right after and before the collision respectively. 

As a consequence, we can contract the momentum conservation (\ref{eq:momcons}) with the KY tensor to see that the sums of vectors associated with KY tensors are always trivially conserved
\begin{align}
& \sum_i (Y_{\mu\kappa} p_{(i)}^{\prime \mu}) - \sum_i (Y_{\mu\kappa} p_{(i)}^{ \mu}) =0\,. \label{eq:colKY}
\end{align}
Similarly, if we contract (\ref{eq:momcons}) with a Killing vector $\xi_\mu$, we see that that the sum of $\xi_\mu p_{(i)}^{ \mu}$ is conserved in collisions. However, as was noticed also by \citet{grantflan2015}, if we try to see what happens to the sum of the quadratic Carter constants $C_{(i)} \equiv K_{\mu\nu} p_{(i)}^{ \mu} p_{(i)}^{ \nu}\,, K_{\mu\nu} \equiv Y_{\mu \kappa} Y^{\kappa}_{\; \nu},\,$ in momentum exchanges, we obtain 
\begin{equation}
\sum_i C_{(i)}' - \sum_i C_{(i)} = \sum_i \sum_{j \neq i} K_{\mu\nu} p_{(i)}^{ \mu} p_{(j)}^{ \nu} - \sum_i \sum_{j \neq i} K_{\mu\nu} p_{(i)}^{\prime \mu} p_{(j)}^{\prime \nu}\,. \label{eq:cartnon}
\end{equation}
By Einstein equivalence principle, {\em every} scattering process will be locally ignorant of the privileged directions of the background space-time and thus also the existence of $K_{\mu\nu}$. Therefore, any scattering process will produce a nonzero right-hand side of (\ref{eq:cartnon}) and will change the value of the sum of Carter constants. 

We will now quickly sketch why these results lead to the fact that collisions will not spoil the divergence-free properties of tensors such as $\rho u^\mu u^\nu \xi_\nu$ or $\rho u^\mu u^\nu Y_{\nu\kappa}$, whereas $\rho u^\mu u^\nu u^\lambda K_{\nu \lambda}$ becomes non-conserved once collisions are included.

Let us assume for simplicity that the collisions are such that particles are not annihilated or created, and that the particle rest mass stays the same at all times. We can then write a distributional or ``skeletonized'' total tensor $\rho u^\nu L^\kappa$ using the worldlines of individual particles $x^\mu_{(i)} (\tau)$ as \citep[see e.g.][]{trautman2002}
\begin{align}
&\rho u^\nu L^\kappa  = \frac{1}{\sqrt{-g}} \sum_i m_i \int_{-\infty}^{\infty} u^\nu_{(i)}  L^{\kappa}_{(i)} \delta^{(4)} \! \left(x^\mu - x^{\mu}_{(i)}(\tau)\right) \d \tau\,, \\
&u^\nu_{(i)} = u^\nu_{(i)}(\tau) =\frac{d x^\nu_{(i)}}{\d \tau} \,,\; L^{\kappa}_{(i)} = L^{\kappa}_{(i)}(\tau) =  \frac{d x^\lambda_{(i)}}{\d \tau} Y_{\lambda}^{\;\,\kappa}\!\left(x^{\mu}_{(i)}(\tau)\right)\,,
\end{align}
where $\rho u^\nu L^\kappa$ obviously integrates out only into a function of $x^\mu$.
Using the identity
\begin{equation}
\frac{\d x^\nu}{\d \tau} \!\left[\delta^{(4)} \! \left(x^\mu - x^{\mu}_{(i)}(\tau)\right)\right]_{,\nu} = - \frac{\d \;}{\d \tau} \delta^{(4)} \!\left(x^\mu - x^{\mu}_{(i)}(\tau)\right)\,,
\end{equation} we then obtain 
\begin{equation}
(\rho u^\nu L^\kappa)_{;\nu} = \frac{1}{\sqrt{-g}} \sum_i m_i \int_{-\infty}^{\infty}  \frac{\mathrm{D} L^{\kappa}_{(i)}}{\d \tau} \delta^{(4)} \! \left(x^\mu - x^{\mu}_{(i)}(\tau)\right) \d \tau \,.
\end{equation}
Now let us recall that during the free-streaming of the particle we have ${\mathrm{D} L^{\kappa}_{(i)}}/{\d \tau} = 0$ and during a collision the vector $L^\kappa_{(i)}$ jumps to another vector $L^{\kappa \prime}_{(i)} = L^\kappa_{(i)} + \Delta^\kappa_{\mathrm{col}(i)}$. We thus have
\begin{equation}
\frac{\mathrm{D} L^{\kappa}_{(i)}}{\d \tau} = \sum_\mathrm{col} \Delta^\kappa_{\mathrm{col}(i)} \delta(\tau - \tau_{\mathrm{col}(i)})\,,
\end{equation}
where $\tau_{\mathrm{col}(i)}$ is the value of the proper time at which the collision happens. The expression for $(\rho u^\nu L^\kappa)_{;\nu}$ then reduces to
\begin{equation}
(\rho u^\nu L^\kappa)_{;\nu} = \frac{1}{\sqrt{-g}}  \sum_\mathrm{col} \delta^{(4)}(x^\mu - x^\mu_\mathrm{col}) \left(\sum_i m_i \Delta^\kappa_{\mathrm{col}(i)} \right)\,, \label{eq:coldiv}
\end{equation}
where $x^\mu_\mathrm{col} = x^\mu_{(i)} (\tau_{\mathrm{col}(i)})$ for every particle $(i)$ taking part in the given collision. We can now see that the sum in the round brackets in (\ref{eq:coldiv}) corresponds to equation (\ref{eq:colKY}) computed at every given collision and we thus obtain
\begin{equation}
(\rho u^\nu L^\kappa)_{;\nu} = 0 \,. \label{eq:KYskoro}
\end{equation}
If we applied this procedure to $\rho u^\mu K_{\kappa \lambda} u^\kappa u^\lambda$, we would obtain $(\rho u^\mu K_{\kappa \lambda} u^\kappa u^\lambda)_{;\mu}$ equal to an expression analogous to the right hand side of (\ref{eq:coldiv}) which would, however, not vanish due to the non-conservation of the sum of Carter constants in collisions.  We can thus see that the current $\rho u^\mu K_{\kappa \lambda} u^\kappa u^\lambda$ conserved for non-collisional dust cannot be generalized in the case of fully general matter fields.

On the other hand, considering that for dust we have $T^{\mu\nu} = \rho u^\mu u^\nu$, we can write $\rho u^\mu L^\kappa = T^{\mu\nu}Y_{\nu}^{\;\,\kappa}$ and the fully general counterpart of (\ref{eq:KYskoro}) can then be alternatively derived using $T^{\mu\nu}_{\;\;\; ;\mu} =0$ and the properties of the KY tensor
\begin{equation}
(T^{\mu\nu}Y_{\nu}^{\;\,\kappa})_{;\mu} = T^{\mu\nu}_{\;\;\;;\mu}Y_{\nu\kappa} + T^{\mu\nu}Y_{\nu\;\;;\mu}^{\;\,\kappa}= 0\,. \label{eq:KYskoroo}
\end{equation}
Hence, equation (\ref{eq:KYskoroo}) could in some sense be understood as the ``hidden conservation law" which follows from the parallel transport of $L_\kappa$ along free test particle motion. Nevertheless, it has the flaw that it is not in the form of a divergence of a vector or a totally antisymmetric tensor, and will thus have no simple integral counterparts. For that reason, we create a current form by a divergence with respect to the dangling index $\kappa$, $j^\mu \equiv (T^{\mu\nu}Y_{\nu}^{\;\,\kappa})_{;\kappa} = T^{\mu\nu;\kappa}Y_{\nu\kappa} $, to obtain
\begin{equation}
\begin{split}
(T^{\mu\nu;\kappa}Y_{\nu\kappa})_{;\mu} &=  (T^{\mu\nu}Y_{\nu}^{\;\,\kappa})_{;\kappa\mu} =  (T^{\mu\nu}Y_{\nu}^{\;\,\kappa})_{;\mu \kappa} \\ &= (T^{\mu\nu}_{\;\;\;;\mu}Y_{\nu\kappa} + T^{\mu\nu}Y_{\nu\;\;;\mu}^{\;\,\kappa})_{;\kappa} = 0 \,,
\end{split}
\end{equation}
where we have used the fact that one can swap the order of divergences in all indices of a tensor. 

That is, we come to the final conclusion that a conserved current linked to the non-trivial conservation laws along geodesics or the ``hidden symmetry'' is $j^\mu = T^{\mu\nu;\kappa}Y_{\nu\kappa}$.

%%%%%%%%%%%%%%%%%%%%%%%%%%%%%%%%%%%%%%%%%%%%%%%%%%%%%%%%%%%%%%%%%%%%%%%%%%%%%%%%%%%%%%%%%%%%%%%%%%%%%%%%%%%%%%%%%%%%%%%%%%%%%%%%%%%%%%%%%%%%%%%%%%%%%%%%%%%%%%%%%

\subsection{Generating tensor for KY current}

One can easily show that the current $j^\mu$ can be (under the assumption of $T^{\mu\nu}_{\;\;\,;\mu} = 0$) also rewritten as a divergence of an antisymmetric tensor $j^\mu= F^{\mu\nu}_{\;\;\; \; ;\nu}$, where
\begin{equation}
 F^{\mu\nu}= 2 T^{\kappa [\mu}Y^{\nu]}_{\;\;\; \kappa}.
\end{equation}
Both $j^\mu_{\;\; ;\mu} =0$ and $j^\mu = F^{\mu\nu}_{\;\;\;\;;\nu}$ are fulfilled due to the properties of the KY tensor and the equations of motion for $T^{\mu\nu}$. Hence, a violation of either is an indication of improper evolution as is shown in more detail in the next Subsection. 

On the other hand, this also points towards a general method of generating a large amount of alternative conserved currents with properties similar to $j^\mu$ either from general tensors or from KY tensors of higher rank; we discuss this possibility in Appendix \ref{app}. It is not clear  what is the usefulness or proper meaning of the whole class of conserved currents and tensors presented in Appendix \ref{app}, but we restrict ourselves here only to the study of the current $j^\mu = T^{\mu\nu;\kappa} Y_{\nu \kappa}$ as a representative member and leave a deeper investigation of the general class as a possibility for future works.

%%%%%%%%%%%%%%%%%%%%%%%%%%%%%%%%%%%%%%%%%%%%%%%%%%%%%%%%%%%%%%%%%%%%%%%%%%%%%%%%%%%%%%%%%%%%%%%%%%%%%%%%%%%%%%%%%%%%%%%%%%%%%%%%%%%%%%%%%%%%%%%%%%%%%%%%%%%%%%%%%

\subsection{Integral forms of KY conservation law} \label{subsec:integral}

Under the assumption  $T^{\mu}_{\;\;\nu;\mu} = r_\nu$ the KY conservation law modifies as $F^{\mu\nu}_{\;\;\;\;;\nu} = j^\mu - r_\alpha Y^{\mu \alpha},\, j^\mu_{\; ;\mu} = (r_\nu Y^{\nu\mu})_{;\mu}$. If we then integrate over the same ranges as in (\ref{eq:expint}), we obtain 
\begin{equation}
\begin{split}
&\left[ \int_V\! j^t  \sqrt{-g} \,\d^3 x\right]_{t_0}^{t_1}  + \int_{t_0}^{t_1}\!\!\!\int_S \! j^i   \sqrt{-g} \,\d^2 S_i \d t =
\\
& \left[ \int_V\! r_\nu Y^{t\nu}  \sqrt{-g} \,\d^3 x\right]_{t_0}^{t_1}  + \int_{t_0}^{t_1}\!\!\! \int_S r_\nu Y^{i\nu}  \sqrt{-g} \,\d^2 S_i  \d t  \,. \label{eq:KYViolInteg}
\end{split}
\end{equation} 
In other words, the KY conservation law is weaker in nature than those coming from explicit symmetries because the dimension of the integration region on the left-hand side of (\ref{eq:KYViolInteg}) is the same as the region in which the residue is detected (the right-hand side).  The reason why it is still worthwhile to consider this conservation law is that it will probe components of $r_\nu$ which are not in the direction of the symmetries of the space-time. Indeed, in the Kerr space-time with non-zero spin, the KY tensor is a  non-degenerate matrix and by a convenient choice of integration intervals we can in principle detect any component of the residue $r_\nu$.

It is also useful to consider the integral form of $F^{\mu\nu}_{\;\;\;\;;\nu} = j^\mu - r_\alpha Y^{\mu \alpha}$ integrated either over a spatial volume $V$ at a fixed time $t=t_0$ or over a spatial surface $S$ from $t_0$ to $t_1$ to yield
\begin{align}
\begin{split}
& \int_S F^{ti}   \sqrt{-g} \,\d^2 S_i\Bigg|_{t=t_0} - \int_V j^t   \sqrt{-g} \,\d^3 x\Bigg|_{t=t_0} =
\\
&  -\int_V\! r_\nu Y^{t\nu}  \sqrt{-g} \,\d^3 x\Bigg|_{t=t_0}
\,, \label{eq:initcheck}
\end{split}
\\
\begin{split}
& \left[ \int_S F^{ti}   \sqrt{-g} \,\d^2 S_i \right]_{t_0}^{t_1}- \int_{t_0}^{t_1}\!\!\!\int_S j^i   \sqrt{-g} \,\d^2 S_i \d t  = 
\\
&   - \int_{t_0}^{t_1}\!\!\! \int_S r_\nu Y^{i\nu}  \sqrt{-g} \,\d^2 S_i  \d t
\,. \label{eq:evolcheck}
\end{split}
\end{align}
The identities (\ref{eq:initcheck}) and (\ref{eq:evolcheck}) can be combined with (\ref{eq:KYViolInteg}) as convenient.

In the following, we will also use the integral form of the ``KY-projected equations of motion" $(T^{\mu\nu}Y_{\nu}^{\;\, \kappa})_{;\mu} = r^\nu Y_{\nu}^{\;\, \kappa}$, which reads
\begin{equation}
\begin{split}
&\left[ \int_V\! T^{t \nu}Y_{\nu}^{\;\, \kappa} \sqrt{-g} \,\d^3 x\right]_{t_0}^{t_1}  + \int_{t_0}^{t_1}\!\!\int_S \! T^{i \nu} Y_{\nu}^{\;\, \kappa}  \sqrt{-g} \,\d^2 S_i \d t 
\\
& + \int_{t_0}^{t_1}\!\!\int_V\! \Gamma^\kappa_{\;\mu \lambda} T^{\mu\nu}Y_{\nu}^{\;\, \lambda} \sqrt{-g} \,\d^3 x
 \, \d t  = \int_{t_0}^{t_1}\!\!\int_V\! r^\nu Y_{\nu\kappa} \sqrt{-g} \,\d^3 x \, \d t \,. \label{eq:quasiKY}
\end{split} 
\end{equation}

%%%%%%%%%%%%%%%%%%%%%%%%%%%%%%%%%%%%%%%%%%%%%%%%%%%%%%%%%%%%%%%%%%%%%%%%%%%%%%%%%%%%%%%%%%%%%%%%%%%%%%%%%%%%%%%%%%%%%%%%%%%%%%%%%%%%%%%%%%%%%%%%%%%%%%%%%%%%%%%%%
%%%%%%%%%%%%%%%%%%%%%%%%%%%%%%%%%%%%%%%%%%%%%%%%%%%%%%%%%%%%%%%%%%%%%%%%%%%%%%%%%%%%%%%%%%%%%%%%%%%%%%%%%%%%%%%%%%%%%%%%%%%%%%%%%%%%%%%%%%%%%%%%%%%%%%%%%%%%%%%%%

\section{Demonstration in Kerr space-time} \label{sec:demo}

The Kerr metric in Boyer-Lindquist coordinates $t,\varphi,r,\vartheta$ reads
\begin{equation}
\begin{split}
\d s^2 =& - \left( 1 - \frac{2Mr}{\Sigma} \right) \d  t^2  + \frac{\Sigma}{\Delta} \d r^2 + \Sigma \d \vartheta^2 \\& + \sin^2 \! \vartheta \left(r^2 + a^2 + \frac{2M r a^2 \sin^2 \! \vartheta}{\Sigma} \right) \d \varphi^2\\& - \frac{4 Mr a \sin^2 \! \vartheta}{\Sigma} \d t \d \varphi\,,
\end{split}
\end{equation}
where $M,a$ are the mass and the spin of the black hole respectively, $\Sigma = r^2 + a^2 \cos^2 \! \vartheta$, and $\Delta = r^2 - 2Mr + a^2$. The components of the Killing-Yano tensor then read \citep{floyd1973}
\begin{equation}
\begin{split}
& Y_{rt} = -Y_{tr} =   a \cos \vartheta \,,\\
& Y_{r\varphi} = -Y_{\varphi r} =  -a^2\cos \vartheta \sin^2 \! \vartheta \,,\\
& Y_{\vartheta \varphi} = -Y_{\varphi \vartheta} = (r^2 + a^2) r \sin \vartheta   \,,\\
&  Y_{\vartheta t} = -Y_{t\vartheta} =  -a r \sin \vartheta \,,\\
& Y_{t\varphi} = - Y_{\varphi t} = Y_{r\vartheta} = - Y_{\vartheta r} = 0 \,. \label{eq:KYcomponents}
\end{split}
\end{equation}
The Killing-Yano tensor is independent of $M$ and it is invertible for $a\neq 0$. An important property is that all the components $Y_{\mu\nu}$ are antisymmetric with respect to a reflection about the equatorial plane $\vartheta \to \mathrm{\pi} - \vartheta$.

%%%%%%%%%%%%%%%%%%%%%%%%%%%%%%%%%%%%%%%%%%%%%%%%%%%%%%%%%%%%%%%%%%%%%%%%%%%%%%%%%%%%%%%%%%%%%%%%%%%%%%%%%%%%%%%%%%%%%%%%%%%%%%%%%%%%%%%%%%%%%%%%%%%%%%%%%%%%%%%%%

\subsection{Dust flow}
Let us now construct an analytical accretion toy model with a parametrized non-physical deviation; this model will be used to study the applicability of the KY conservation law in the next Subsection. We set up a field of dust $T^{\mu\nu} = \rho u^\mu u^\nu$ which is infalling into a Kerr black hole with a mass density $\rho$ and the four-velocity
\begin{align}
& u_\varphi=u_\vartheta=0\,, \\
& u_t = -1 \,, \\
& u_r = -\frac{ \sqrt{2Mr(r^2 + a^2)}}{\Delta} +  \varepsilon\,,
\end{align}
where when the dimensionless constant $ \varepsilon = 0$ then the velocity field fulfils the geodesic equation $a^\mu =0$ but violates it for $ \varepsilon \neq 0$ by introducing two non-zero components of acceleration
\begin{align}
a_\vartheta = & - \varepsilon\frac{ 2  a^2 \sin \vartheta \cos \vartheta \sqrt{2Mr(r^2 + a^2)}}{\Sigma^2}  + \mathcal{O}( \varepsilon^2)  \,, 
\\
a_r = & -  \varepsilon\left(\frac{ \sqrt{2Mr(r^2 + a^2)}}{\Sigma}\right)_{,r}  + \mathcal{O}( \varepsilon^2) \,.
\end{align}
Since the velocity field is purely radial, the continuity equation $(\rho u^\mu)_{;\mu} =0$ has the simple solution
\begin{equation}
\rho = -\frac{\dot{M}(\vartheta)}{2\uppi \Delta u_r} \,, \label{eq:rho}
\end{equation}
where $\dot{M}(\vartheta)$ is an arbitrary positive function of $\vartheta$  which represents the dust accretion rate through a $\vartheta=\mathrm{const.}$ layer. 

This flow is parametrized by the arbitrary function $\dot{M}(\vartheta)$ and the deviation parameter $ \varepsilon$. Since the continuity equation is fulfilled, we obtain that the residue vector is $T^{\mu}_{\;\;\nu;\mu} =r_{\nu} = \rho a_\nu$.

%%%%%%%%%%%%%%%%%%%%%%%%%%%%%%%%%%%%%%%%%%%%%%%%%%%%%%%%%%%%%%%%%%%%%%%%%%%%%%%%%%%%%%%%%%%%%%%%%%%%%%%%%%%%%%%%%%%%%%%%%%%%%%%%%%%%%%%%%%%%%%%%%%%%%%%%%%%%%%%%%

\subsection{Integral test} \label{subsec:test}
The three ``usual" conservation laws we would check in this situation are conservation of mass $(\rho u^\mu)_{;\mu} =0$, conservation of angular momentum $(T^{\mu}_{\;\;\varphi})_{ ;\mu} = 0$, and conservation of energy $(T^{\mu}_{\;\;t})_{ ;\mu} = 0$. However, it is easy to verify that the example has been constructed so as to automatically fulfil these conservation laws while violating the evolution equations in the ``blind spots" of these tests. 

We will now compare the detection of a non-zero $ \varepsilon$ by the equalities (\ref{eq:quasiviol}), which was obtained from a direct integration of the equations of motion, (\ref{eq:quasiKY}), which was obtained from the KY-projected equations of motion, and (\ref{eq:initcheck}), which was obtained from the KY conservation law. 

As for the implementation of (\ref{eq:quasiviol}), we choose the $\nu=r$ component for our purposes, take the volume $V$ as the volume between $r_0$ and $r_1$, and obtain under the assumption of automatic stationarity and axisymmetry
\begin{align}
\begin{split}
&\left[ \int_0^\uppi\!\!  \rho (u_r)^2 \Delta \sin \vartheta\, \d\vartheta \right]_{r_0}^{r_1} \\
& -  \frac{1}{2}\int_{r_0}^{r_1}\!\!\!\! \int_0^\uppi \!\! g_{\alpha\beta,r} \rho u^\alpha u^\beta \Sigma  \sin \vartheta \, \d \vartheta \, \d r \\
& = \int_{r_0}^{r_1}\!\!\!\! \int_0^\uppi \!\! \rho a_r \, \Delta  \sin \vartheta \, \d \vartheta \, \d r\,. \label{eq:Tmunurtest}
\end{split} 
\end{align}
To implement (\ref{eq:quasiKY}), we must choose a component which detects the non-zero acceleration in the meridional plane. A quick glance at the non-zero components of the KY tensor (\ref{eq:KYcomponents}) shows that we can detect the meridional acceleration only through the $\kappa=t,\varphi$ components of (\ref{eq:quasiKY}). Here we choose the $\kappa=t$ component and obtain similarly to (\ref{eq:Tmunurtest})
\begin{equation}
\begin{split}
&\left[ \int_0^\uppi\!\!  \rho u_r L^t \Delta \sin \vartheta\, \d\vartheta \right]_{r_0}^{r_1} \\
& +  \frac{1}{2}\int_{r_0}^{r_1}\!\!\!\! \int_0^\uppi \!\! \Gamma^{t}_{\; \mu \lambda} \rho u^\mu L^\lambda \Sigma  \sin \vartheta \, \d \vartheta \, \d r \\
& = \int_{r_0}^{r_1}\!\!\!\! \int_0^\uppi \!\! \rho (a^r Y_{r}^{\;t} + a^\vartheta Y_{\vartheta}^{\;t})\, \Sigma  \sin \vartheta \, \d \vartheta \, \d r\,. \label{eq:TYrtest}
\end{split} 
\end{equation}
Finally, we implement (\ref{eq:initcheck}) with the volume between $r=r_0$ and $r=r_1$ in place of $V$ to obtain
\begin{equation}
\begin{split}
&\left[ \int_0^\uppi\!\!  F^{tr}  \Sigma \sin \vartheta\, \d\vartheta \right]_{r_0}^{r_1} \\
& -  \int_{r_0}^{r_1}\!\!\!\! \int_0^\uppi \!\! j^t  \Sigma  \sin \vartheta \, \d \vartheta \, \d r \\
& = \int_{r_0}^{r_1}\!\!\!\! \int_0^\uppi \!\! \rho (a^r Y_{r}^{\;t} + a^\vartheta Y_{\vartheta}^{\;t})\, \Sigma  \sin \vartheta \, \d \vartheta \, \d r\,. \label{eq:KYrtest}
\end{split}
\end{equation}
In all the cases, we have to exclude the black hole from the integration region because its central singularity acts as a sink where conservation laws are violated. 

The integrals on the left-hand sides of (\ref{eq:Tmunurtest}), (\ref{eq:TYrtest}), and (\ref{eq:KYrtest}) should in every case detect deviations (non-zero right-hand sides) which are of order $\mathcal{O}( \varepsilon )$ and correspond to quantities integrated over the same coordinate ranges. Furthermore, the integration regions and components in (\ref{eq:TYrtest}) and (\ref{eq:KYrtest}) were chosen such that they have the identical projections of $r_\nu$ on the right-hand sides. 

There is one very important difference between (\ref{eq:TYrtest}) and (\ref{eq:KYrtest}) as compared to (\ref{eq:Tmunurtest}), and that is the response of these integrals to terms symmetric or antisymmetric with respect to reflection about the equatorial plane. Due to the reflection antisymmetry of $Y_{\mu\nu}$ and the reflectionally symmetric integration region, the KY-based expressions (\ref{eq:TYrtest}) and (\ref{eq:KYrtest}) measure the reflectionally antisymmetric part of $\rho a^\mu$, and (\ref{eq:Tmunurtest}) the symmetric part. Since we have constructed $a^\mu$ as reflectionally symmetric, we need to introduce reflection asymmetry in $\rho$ for the KY tests (\ref{eq:TYrtest}) and (\ref{eq:KYrtest}) to yield non-trivial results. We do so by choosing the angular mass influx from (\ref{eq:rho}) as $\dot{M}(\vartheta) = 1 + \cos \vartheta$.

The numerical demonstration of these tests now consists of a detection of a non-zero $ \varepsilon$ when the dust stress-energy tensor is given to us in terms of double-precision numbers on a finite grid. The grid is constructed in Boyer-Lindquist coordinates $r,\vartheta$ with spacing $\delta r = h,\,\delta \vartheta = h/r$ where $h$ is some given length constant. The left-hand sides of (\ref{eq:Tmunurtest}), (\ref{eq:TYrtest}), and (\ref{eq:KYrtest}) are then evaluated using numerical integration. To evaluate the integral of $j^t $ we must, however, compute the gradients of the stress-energy tensor by numerical differentiation and only after that perform numerical integration on the finite grid. 

We numerically approximate the gradient by the symmetric difference $f'(x) = [f(x+h)- f(x-h)]/2h + \mathcal{O}(h^3)$ and the integrals via the trapezoidal rule $\int f(x) \d x = h (f_0/2 + f_1 + ... + f_{n-1} + f_n/2) + \mathcal{O}(h^2)$ in given coordinates. We perform all the computations at double precision and define the ``base-noise'' $\sigma_0$ as the numerically computed value of the left-hand sides at $ \varepsilon =0$. We then define the detection value $ \varepsilon_\mathrm{d}$ of $ \varepsilon$ as the value for which the left-hand evaluates as $10 \sigma_0$.

%%%%%%%%%%%%%%%%%%%%%%%%%%%%%%%%%%%%%%%%%%%%%%%%%%%%%%%%%%%%%%%%%%%%%%%%%%%%%%%%%%%%%%%%%%%%%%%%%%%%%%%%%%%%%%%%%%%%%%%%%%%%%%%%%%%%%%%%%%%%%%%%%%%%%%%%%%%%%%%%%

\subsection{Results}
\begin{figure}
\begin{center}
\includegraphics[width=0.48\textwidth]{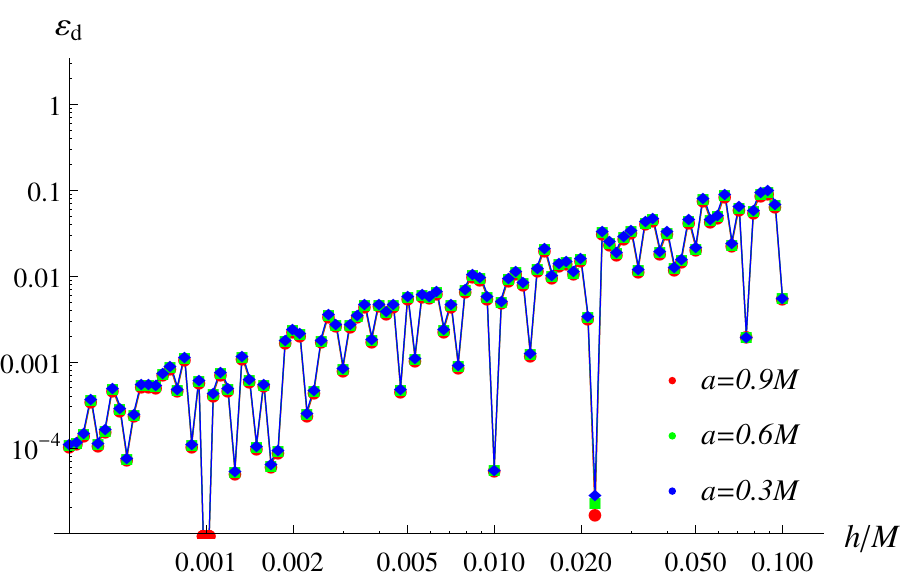}
\caption{The detection value of the non-physical perturbation $\varepsilon_\mathrm{d}$ as detected by the integrated equations of motion (\ref{eq:Tmunurtest}) as a function of grid spacing $h$. The integration bounds $[r_0,r_1]$ are taken as $[3M,6M]$ and various values of spin were used.} \label{fig:Ttest}
\end{center}
\end{figure}

\begin{figure}
\begin{center}
\includegraphics[width=0.48\textwidth]{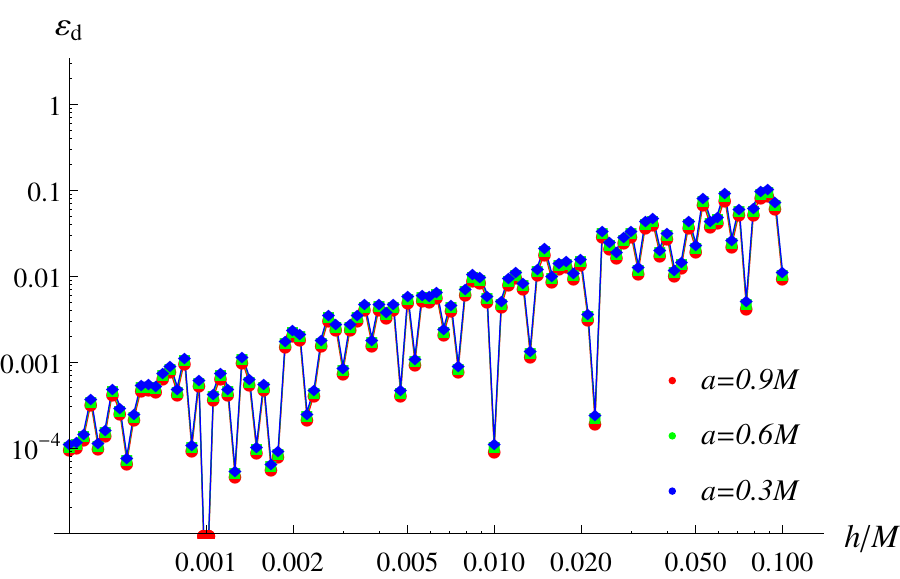}
\caption{The detection value of the non-physical perturbation $\varepsilon_\mathrm{d}$ as detected by the KY-projected equations of motion (\ref{eq:TYrtest}) as a function of grid spacing $h$. Same integration bounds and values of spin as in Fig. \ref{fig:Ttest} are used and the scaling and ranges of axes are also identical to Fig. \ref{fig:Ttest}.} \label{fig:TYtest}
\end{center}
\end{figure}

\begin{figure}
\begin{center}
\includegraphics[width=0.48\textwidth]{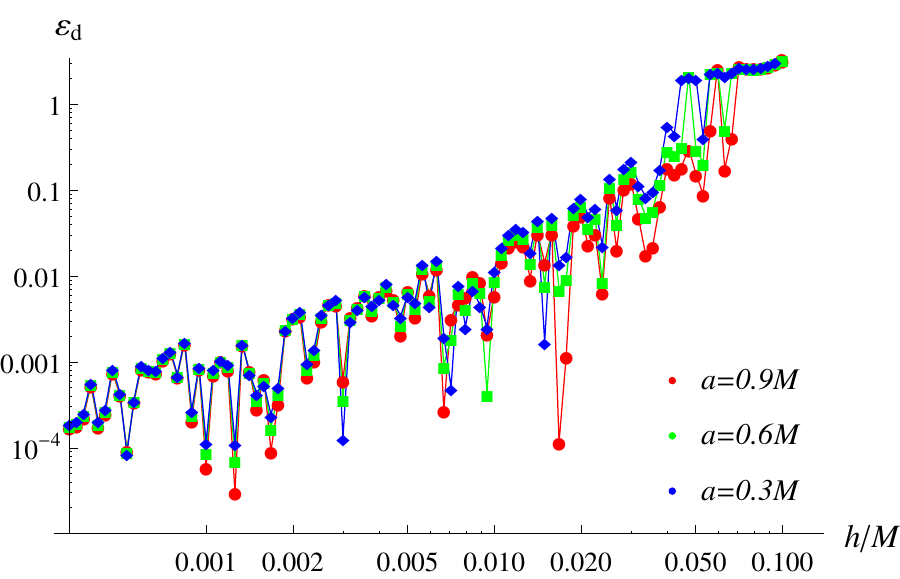}
\caption{The detection value of the non-physical perturbation $\varepsilon_\mathrm{d}$ as detected by the KY conservation law (\ref{eq:KYrtest}) as a function of grid spacing $h$. Same integration bounds and values of spin as in Fig. \ref{fig:Ttest} are used and the scaling and ranges of axes are also identical to Fig. \ref{fig:Ttest}.} \label{fig:KYtest}
\end{center}
\end{figure}

In Figures \ref{fig:Ttest},\ref{fig:TYtest}, and \ref{fig:KYtest} we plot the dependence of $ \varepsilon_\mathrm{d}$ on $h$ for various values of the spin parameter $a$ for (\ref{eq:Tmunurtest}), (\ref{eq:TYrtest}), and (\ref{eq:KYrtest}) respectively.

%---------------------------------------------------------------------------------------------------------------------------------------------------%
%---------------------------------------------------------------------------------------------------------------------------------------------------%

%---------------------------------------------------------------------------------------------------------------------------------------------------%
%---------------------------------------------------------------------------------------------------------------------------------------------------%

%---------------------------------------------------------------------------------------------------------------------------------------------------%
%---------------------------------------------------------------------------------------------------------------------------------------------------%

The results are largely independent of $a$ since the relative sizes of the terms in (\ref{eq:Tmunurtest}), (\ref{eq:TYrtest}), and (\ref{eq:KYrtest}) stay constant, to linear order in $a$, independently of the value of spin. When $\varepsilon = 1$, then the deviation of our model velocity from the physical one is of orders of the speed of light, so it is reasonable to require that the numerical test definitely detects $\varepsilon \lesssim 1$. We see in Fig. \ref{fig:KYtest} that the test based on the KY conservation law is noisier due to numerical differentiation and requires $h \lesssim 0.05M$ to detect perturbations within $\varepsilon \lesssim 1$. On the other hand, the tests (\ref{eq:Tmunurtest}) and (\ref{eq:TYrtest}) are free of the noise coming from numerical differentiation and detect $\varepsilon \lesssim 1$ already for much larger grid spacings. Nonetheless, once the step is $h \lesssim 0.05M$, the KY conservation law yields comparable results to the (\ref{eq:Tmunurtest}) and (\ref{eq:TYrtest}).   

\section{Discussion and concluding remarks}

Even though checking for the conservation of the KY current $j^\mu = T^{\mu\nu;\kappa} Y_{\nu \kappa}$ can detect violations of the equations of motion which are not detectable by checking for the conservation of particle number, energy, or azimuthal angular momentum, it does so at a requirement of rather fine grid spacing ($h \lesssim 0.05M$) and at a testing power ``only'' comparable to checking the integrated version of the equations of motion $T^{\mu\nu}_{\;\;\; ;\mu} = 0$ or the ``KY-projected equations of motion'' $(T^{\mu\nu} Y_\nu^{\;\, \kappa})_{;\mu} = 0$. 

In our example, the KY conservation law as well as the KY projected equations of motion provided information about the reflectionally antisymmetric part of the non-physical perturbation while the integrated equation of motion provided information about the symmetric part. Since it will be a general pattern that the KY tests provide information independent of the one obtained by the plain coordinate form of the equations of motion, it is advisable to use at least one of the two in a ``testing toolkit" when investigating the validity of numerical evolutions. Clearly, the integrated version of KY-projected equations of motion is preferable amongst the two because it does not require the evaluation of either spatial or temporal gradients of matter variables and thus avoids any difficulties with implementation or excessive requirements on grid spacing.

Nevertheless, we must admit that the exploitation of the hidden symmetry bears less powerful results than we hoped for. A KY conservation law is possible to formulate and link to the properties of geodesic motion, but it turns out to be only a weak one and, albeit a possibly privileged specimen, a member of a larger class of similarly weak conservation laws which have no longer anything to do with the hidden symmetry. In principle, even the KY-projected equation of motion $(T^{\mu\nu} Y_\nu^{\;\, \kappa})_{;\mu} = 0$ can be considered only as a simpler and more elegant member of a general class of identities $(T^{\mu\nu} X_\nu^{\;\,\kappa})_{;\mu} - T^{\mu\nu}X_{\nu\;\;\;;\mu}^{\;\,\kappa} =0$ for arbitrary $X_{\mu\nu}$. On the other hand, the fact that our results are connected to more general mathematical structures does not lessen any of the statements given in the paragraphs above. 

In upcoming works, we plan to use the hidden symmetry of the Kerr space-time in a different manner. Specifically, we plan to investigate analytical solutions of fluid flows, already hinted upon in \citet{harisconservation}, that share or generalize the hidden symmetry of the Kerr background.

\section*{Acknowledgements}
I would like to thank Charalampos Markakis, Volker Perlick, and Pavel Jefremov for useful discussions on this topic. I am also grateful for the support from a Ph.D. grant of the German Research Foundation within its Research Training Group 1620 {\em Models of Gravity}. 

%%%%%%%%%%%%%%%%%%%%%%%%%%%%%%%%%%%%%%%%%%%%%%%%%%



\bibliographystyle{mnras}
\bibliography{literatura}

%%%%%%%%%%%%%%%%%%%%%%%%%%%%%%%%%%%%%%%%%%%%%%%%%%

%%%%%%%%%%%%%%%%% APPENDICES %%%%%%%%%%%%%%%%%%%%%
\appendix
\section{Conservation laws from general tensors} \label{app}
Consider an arbitrary tensor $X^{\mu\nu}$ and a tensor analogous to $F^{\mu\nu}$
\begin{equation}
\tilde{F}^{\mu\nu} = 2 T^{\kappa [\mu} X^{\nu]}_{\;\;\kappa}\,.
\end{equation}
We then see that independent of the properties of $X^{\mu\nu}$, the following current will be conserved
\begin{equation}
\tilde{j}^\mu = \tilde{F}^{\mu\nu}_{\;\;\; ;\nu} =  T^{\mu \nu;\kappa} X_{\kappa \nu} + T^{\kappa \mu} X^\nu_{\;\, \kappa;\nu} - T^{\kappa \nu} X^\mu_{\;\, \kappa;\nu}\,. \label{eq:pseudocons}
\end{equation}
As for the testing power of $\tilde{j}^\mu_{\;;\mu} = 0$ we obtain analogously to (\ref{eq:KYViolInteg}) the detection of $T^{\mu\nu}_{\;\;\;;\mu} = r^\nu$ as
\begin{equation}
\begin{split}
&\left[ \int_V\! \tilde{j}^t  \sqrt{-g} \,\d^3 x\right]_{t_0}^{t_1}  + \int_{t_0}^{t_1}\!\!\!\int_S \! \tilde{j}^i   \sqrt{-g} \,\d^2 S_i \d t =
\\
& \left[ \int_V\! r_\nu X^{t\nu}  \sqrt{-g} \,\d^3 x\right]_{t_0}^{t_1}  + \int_{t_0}^{t_1}\!\!\! \int_S r_\nu X^{i\nu}  \sqrt{-g} \,\d^2 S_i  \d t  \,. \label{eq:pseudoViolInteg}
\end{split}
\end{equation} 
If we investigate under which conditions the last two terms in (\ref{eq:pseudocons}) vanish so that a form $\tilde{j}^\mu = T^{\mu \nu;\kappa} X_{\kappa \nu}$ similar to the KY conservation law is obtained, we get
\begin{equation}
X_{\mu(\nu;\kappa)} =  g_{\mu(\nu} X^\lambda_{\;\, \kappa);\lambda} \,. \label{eq:confKY}
\end{equation}
If $X^{\mu\nu}$ were additionally antisymmetric, then the fulfilment of condition (\ref{eq:confKY}) would make it a so-called conformal KY tensor (of which the usual KY tensors are a subclass with $X^\lambda_{\;\, \kappa;\lambda} = 0$). This suggest that the currents $\tilde{j}^\mu$ derived from KY or conformal KY tensors may play a privileged role as compared to those derived from general tensors. A full investigation of whether this is true is out of the scope of the current paper.

Similarly, it is possible to generate conserved anti-symmetric tensors $\tilde{j}^{\mu_1 ... \mu_{n-1}},\, \tilde{j}^{\mu_1 ... \mu_{n-1}}_{\;\;\;\;\;\;\;\;\;\;\;\;\; ;\mu_{n-1}} = 0$ from any tensor $X^{\mu_1... \mu_n}$. We show the resulting formulas only for $\tilde{j}^{\mu_1 ... \mu_{n-1}}$ generated from KY tensors. 

A KY tensor of arbitrary rank is defined as a totally antisymmetric tensor $Y_{\mu\nu...\kappa} = Y_{[\mu\nu...\kappa]}$ whose gradient is also antisymmetric  $Y_{\mu\nu...\kappa;\gamma} = -Y_{\mu\nu...\gamma;\kappa}$. The tensor analogous to $F^{\mu\nu}$ is then, for a KY tensor of rank $n$, defined as
\begin{equation}
\begin{split}
F^{\mu_1 ... \mu_n} & = \frac{n}{n-1} Y_{\alpha}^{\;\;\; [\mu_1... \mu_{n-1}} T^{\mu_n] \alpha} 
\end{split}
\end{equation}
Using the properties of the Killing-Yano tensor and $T^{\mu\nu}_{\;\;\; ;\nu} = 0$ we then obtain
\begin{equation}
j^{\mu_1... \mu_{n-1}} = F^{\mu_1 ... \mu_n}_{\;\;\;\;\;\;\;\;\;\;\; ;\mu_n} = Y_{\alpha \beta}^{\;\;\; [\mu_1... \mu_{n-2}} T^{\mu_{n-1}] \alpha;\beta} \,.
\end{equation}
It is easy to see that by construction $j^{\mu_1 ... \mu_{n-1}}_{\;\;\;\;\;\;\;\;\;\;\;\;\; ;\mu_{n-1}} = 0$. Since the tensors $F^{\mu_1 ... \mu_n} $ and $j^{\mu_1 ... \mu_{n-1}}$ are totally antisymmetric, integral formulas analogous to the ones in Subsection \ref{subsec:integral} will apply to them. 

Some astrophysically relevant space-times admit KY tensors of higher rank \citep{howarth2000} but these generate redundant conserved quantities for free test particle motion and are thus expected to generate linearly dependent variants of the conservation laws presented in the main text. Of broader theoretical interest is the fact that non-redundant KY tensors of higher rank arise for higher-dimensional spinning black holes \citep{krtous2007} and there the formulas above will probably give independent conservation laws.

%%%%%%%%%%%%%%%%%%%%%%%%%%%%%%%%%%%%%%%%%%%%%%%%%%

% Don't change these lines
\bsp	% typesetting comment
\label{lastpage}
\end{document}